\documentclass[3p,times,procedia]{elsarticle}
\usepackage{nupha_ecrc}

\volume{00}

\firstpage{1}

\journalname{Nuclear Physics A}

\jid{nupha}

\jnltitlelogo{Nuclear Physics A}

\usepackage{amssymb}
\usepackage{amsmath}

\usepackage[figuresright]{rotating}

\begin{document}

\begin{frontmatter}



\dochead{XXVIIth International Conference on Ultrarelativistic Nucleus-Nucleus Collisions\\ (Quark Matter 2018)}

\title{A Quasiparticle Transport Explanation for Collectivity in the Smallest of Collision Systems ($p+p$ and $e^{+}e^{-}$)}


\author{J.L. Nagle}
\address{University of Colorado Boulder and CEA/IPhT/Saclay}
\author{J. Orjuela Koop}
\address{University of Colorado Boulder}


\begin{abstract}
The field of heavy ion physics is at a crossroads in understanding experimental signatures of collectivity in small collision systems, $p+p$ and $p(d,^3\text{He})$+A, at RHIC and the LHC.   A wealth of data obtained in the latter class of asymmetric systems indicate the existence of particle emission patterns similar to those observed in larger A+A collisions~\cite{Nagle:2018nvi},  raising the question of whether the same physics is at play in both cases, lest the cruelty of nature be somehow exposed.   In this talk, we present an extension of earlier studies using the quasiparticle transport model \textsc{ampt} to predict particle emission patterns in the smallest of collision systems, namely $p+p$ and $e^{+}e^{-}$.   The $e^{+}e^{-}$ results have been previously published~\cite{Nagle:2017sjv} and we thus focus here on an extended set of calculations, as shown at the Quark Matter 2018 Conference.
\end{abstract}

\end{frontmatter}


\section{Quasiparticle Transport Studies}
In order to understand the basic dynamics underlying small and large nuclear collisions, it is 
imperative to explore different theoretical approaches and to push them beyond their usual confines of applicability.   In 
that vein, one can model the medium formed in nuclear collisions as a collection of well-defined quasiparticles which scatter according to the Boltzmann transport formalism.   To precisely establish what is meant by ``scattering quasiparticles'', we consider three characteristic length scales: $(i)$ the mean free path of particles $\lambda_{\text{mfp}}$; $(ii)$ the typical
inter-particle spacing $\ell$, and $(iii)$ the formation length of particles $\ell_{form}=\tau_{form} \times v$, corresponding to the time after which they become ``well-defined'' in the sense of admitting a semi-classical description.   A-Multi-Phase-Transport Model (\textsc{ampt})~\cite{Lin:2004en} is a publicly available kinetic transport code which solves the Boltzmann equation numerically for $2\rightarrow 2$ parton scattering after they have traversed some initial formation length.
Calculations using this tool have appeared in a multitude of papers over the years, and have been compared with collectivity observables measured at RHIC and the LHC in both large and small systems. It has been found that, in \textsc{ampt}, it is possible to develop large flow coefficients $v_n$ even in small systems where the partons experience very few scatters, owing to the so-called anisotropic \textit{escape mechanism}~\cite{He:2015hfa}. In these proceedings, we extend these studies to the even smaller $p+p$ and $e^+e^-$ collisions.

\begin{figure*}[hbtp]
\centering
\includegraphics[width=0.4\linewidth]{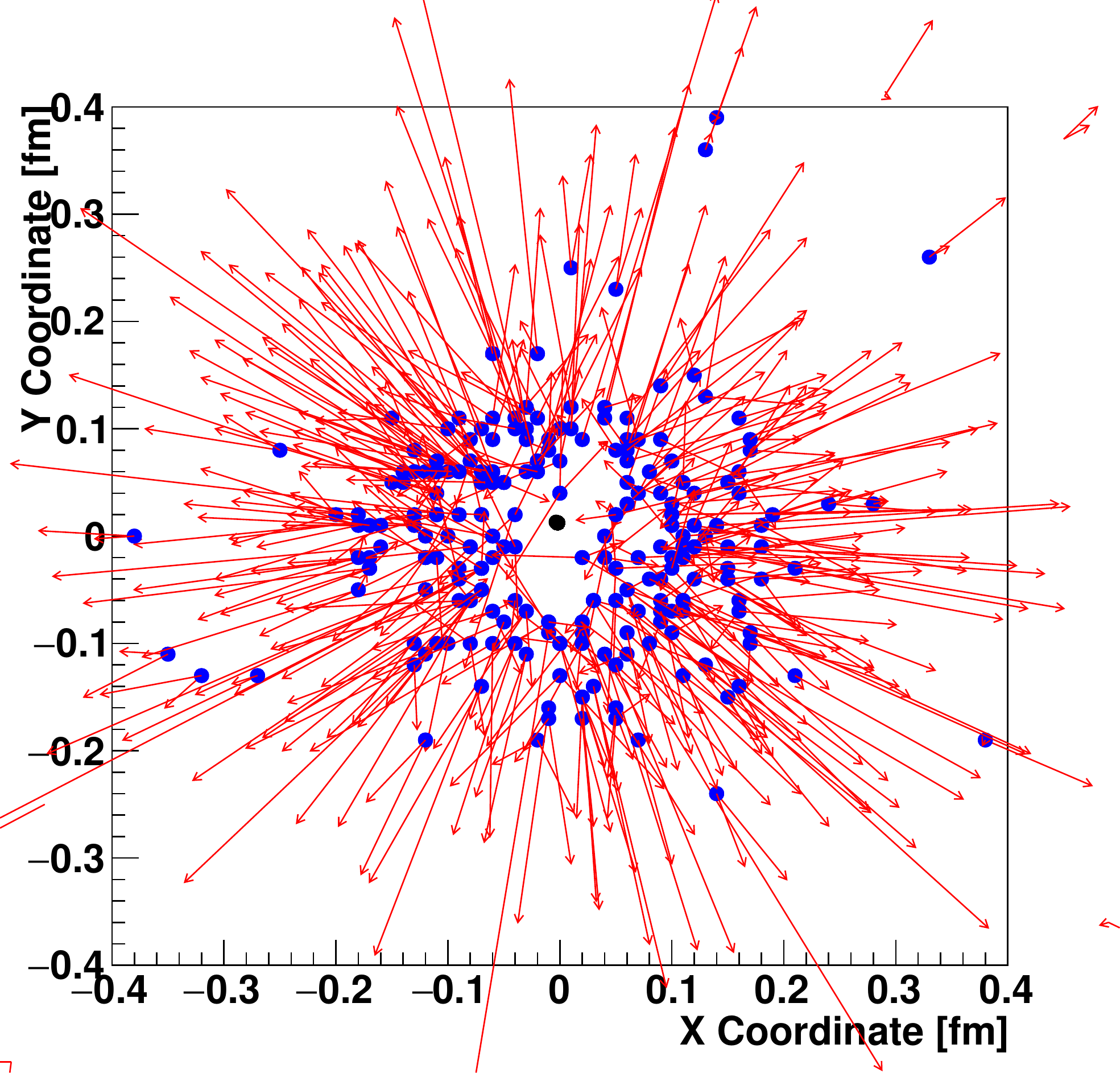}
\includegraphics[width=0.4\linewidth]{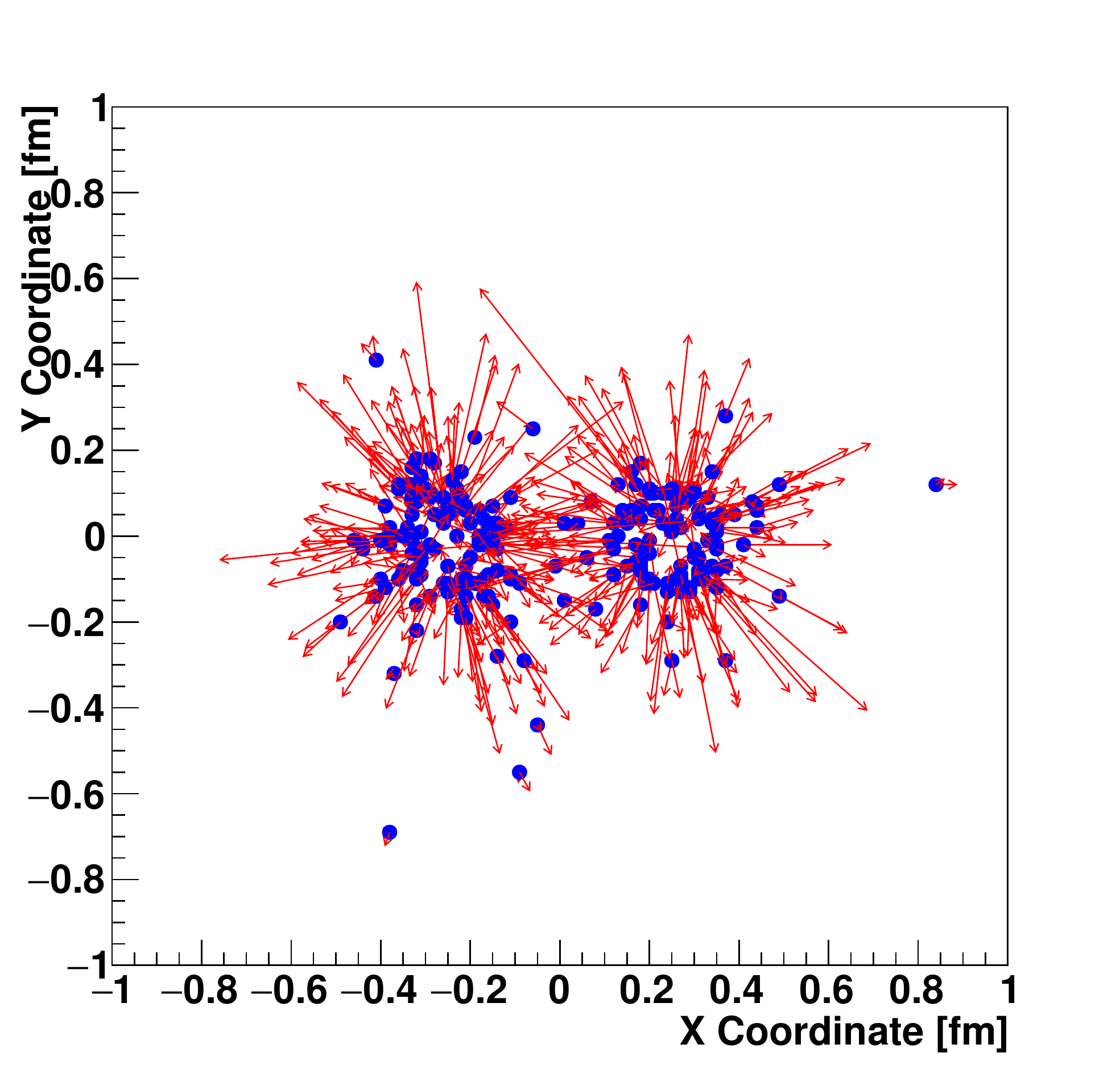}
\caption{\textsc{ampt} event displays showing the initial parton positions and momentum vectors in $e^{+}e^{-}$ (left) and artificially constructed
two string (right) cases.  Note the different x,y axis ranges between the left and right panels.}
\label{fig:stringdisplay}
\end{figure*}

In a modified version of \textsc{ampt}, we simulate $e^{+}e^{-}$ collisions at an energy equal to the $Z$ boson mass via the
generation of a single string extended between a quark and anti-quark pair.   
Figure~\ref{fig:stringdisplay} (left panel) shows the transverse coordinates of partons tunneling from this single string in a high-multiplicity event.   It is notable that the string has a point-like transverse 
extent and that the parton initial positions being at $r \approx 0.1$~fm and with a radially outward
initial velocity vector is simply due to \textsc{AMPT} treating the partons as free-streaming up to their
formation length $\ell_{form}$.   As published in Ref.~\cite{Nagle:2017sjv}, although a fraction of these
partons do undergo scattering, there is little room for significantly redirecting the partons even in events where the random initial orientation of the partons has some particular non-circular geometry.

In the default version of \textsc{ampt}, $p+p$ collisions result in two color strings extended between the protons, separated by a distance equal to their impact parameter.   As a control test, we generate events with two color strings separated by a constant distance of 0.5~fm as shown in the right panel of Figure~\ref{fig:stringdisplay}.   As detailed in Ref.~\cite{Nagle:2018nvi}, this two-string configuration leads to significant final state $v_{2}$, despite the fraction of partons that scatter being significantly less than unity.  We highlight that in \textsc{ampt} when partonic cross sections are small (e.g., $0.75-3.0$ mb), the distribution of the angular separation between pairs of scattered partons is quite isotropic (i.e., not forward peaked). Thus, the scattering between partons moving in opposite directions along the horizontal coordinate will redirect many partons in the perpendicular direction, thus leading to azimuthal anisotropy.

\begin{figure*}[hbtp]
\centering
\includegraphics[width=0.9\linewidth]{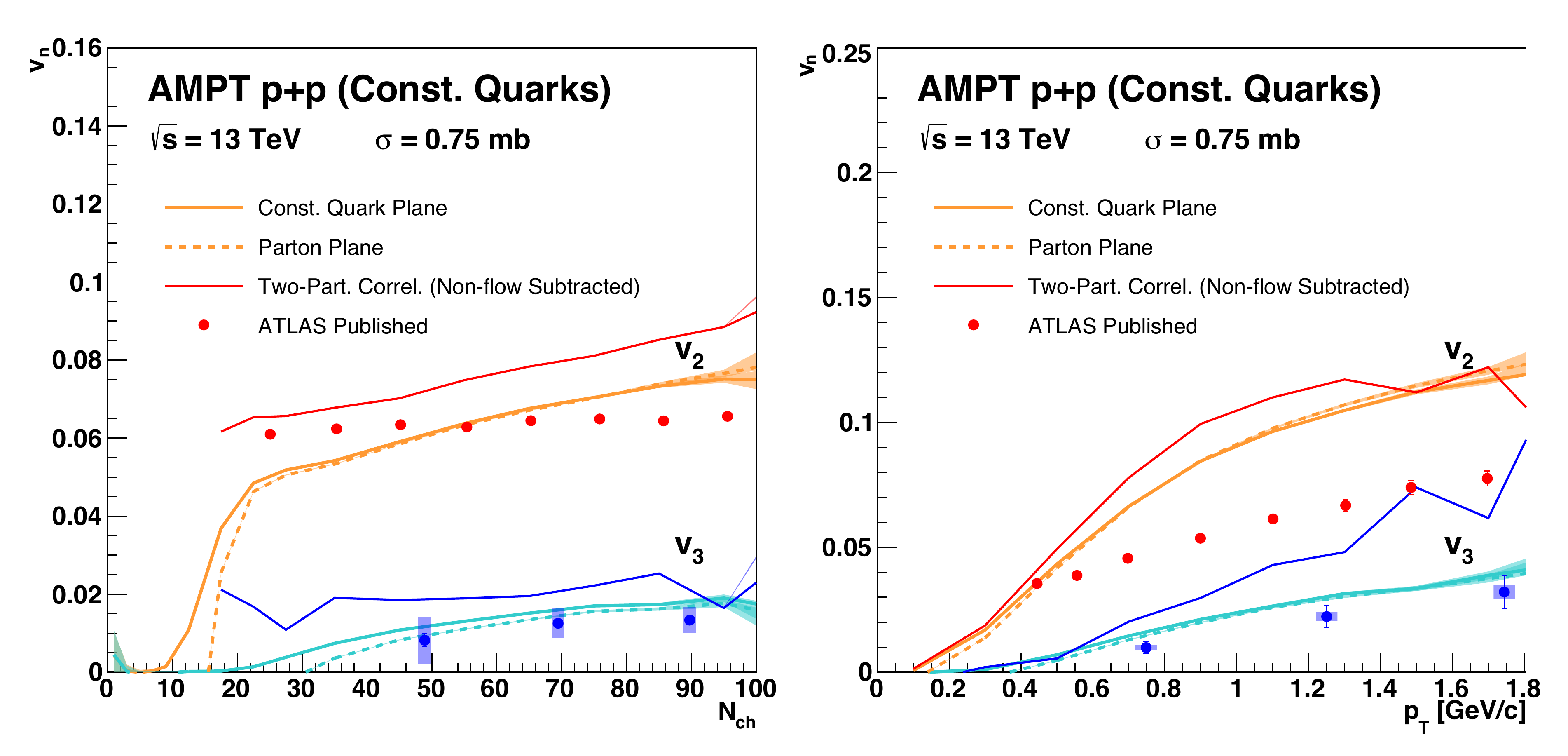}
\caption{ATLAS data for p+p collisions at 13 TeV flow coefficients using the template fitting method.   Also shown are three calculations using AMPT including following the full ATLAS procedure.}
\label{fig:v2_pp_mult}
\end{figure*}

Following on this control test with two strings, we implemented a constituent quark framework in \textsc{ampt} to model the intrinsic geometry of $p+p$ collisions. In this manner, each proton is composed of three constituent quarks that can be thought of as being surrounded by a gluon cloud. Thus, rather than colliding nucleons, the initial Monte Carlo Glauber stage of \textsc{ampt} provides a model of collisions among constituent quarks with an inelastic cross section chosen such that the overall proton-proton cross section matches the experimentally measured value. Since the \textsc{ampt} event record contains all produced final-state hadrons, it is possible to implement the identical methods that have been used to experimentally measure $v_2$ and $v_3$ in $p+p$ at the LHC.
Figure~\ref{fig:v2_pp_mult} shows such results from the ATLAS experiment~\cite{Aaboud:2016yar} as a function of charged particle event multiplicity (left) and---for high multiplicity events---transverse momentum (right). We compare these results to $v_n$ obtained with  \textsc{ampt} using two different methods: $(i)$ the coefficients are computed relative to the \textit{true} initial geometry---as determined from participant nucleons or early-stage partons---and $(ii)$ they are calculated following the ATLAS template fitting method for non-flow subtraction. The latter result provides for the most direct comparison with experimental data. As shown in the figure, \textsc{ampt} appears to capture the qualitative features of the data, while overpredicting $v_{n}$ at high values of $p_{T}$.  It is important to note that the $p_{T}$ dependence of $v_n$ in \textsc{ampt} arises from the momentum-dependent
parton formation length $\ell_{form}$, as well as the hadron coalescence mechanism.

\begin{figure*}[hbtp]
\centering
\includegraphics[width=0.4\linewidth]{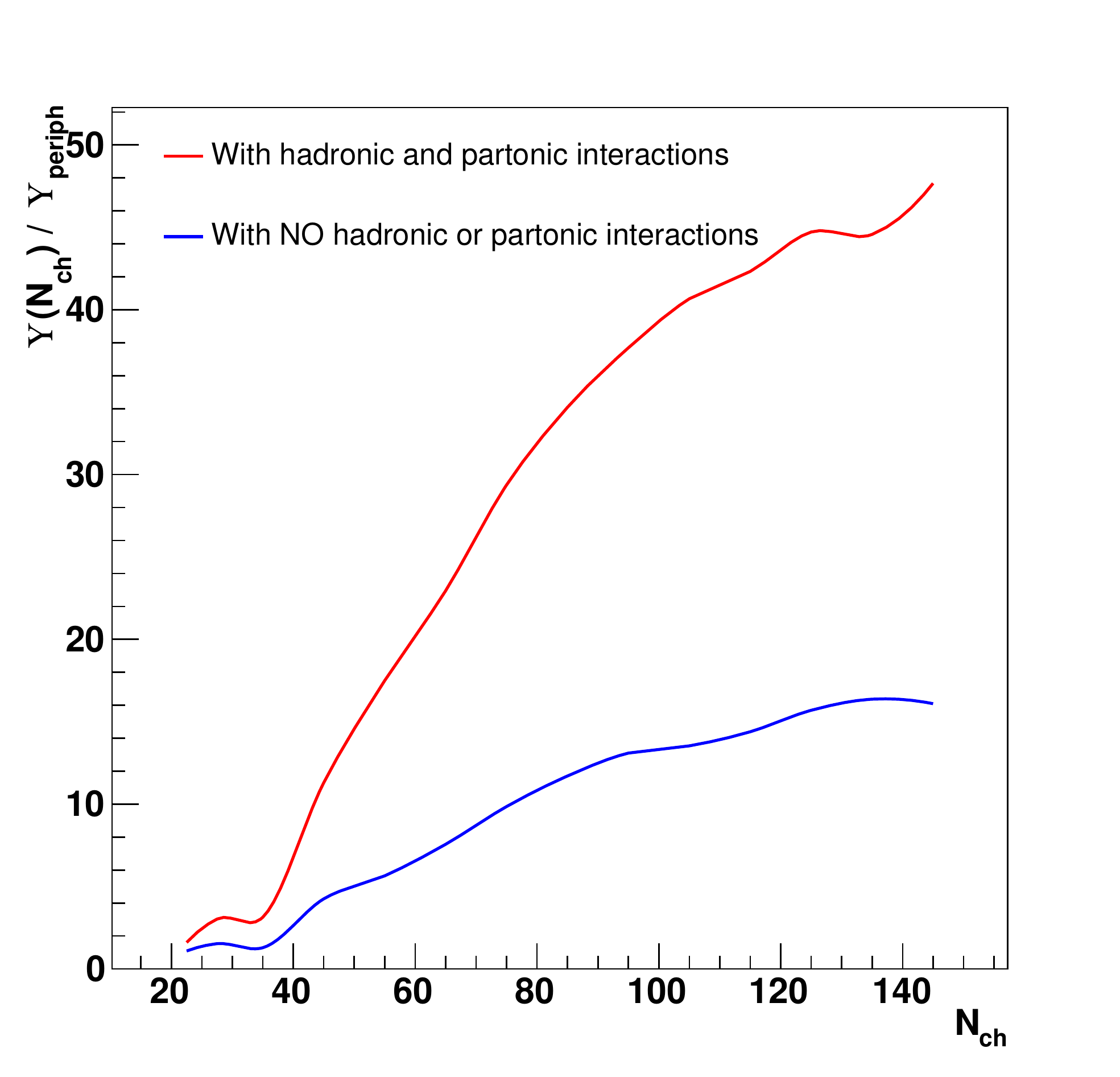}
\caption{Shown are the near-side
jet yield from \textsc{ampt} as a function of event multiplicity for two cases:  with and without hadronic and partonic scattering turned on.}
\label{fig:jetyield}
\end{figure*}

It is of interest that the CMS experiment has also measured $v_n$ in $p+p$ collisions at $\sqrt{s}=13$ TeV, with results that tend towards zero at low $N_{ch}$~\cite{Khachatryan:2016txc}, in marked contrast to the previously discussed ATLAS measurements. This stems from a different non-flow subtraction method that uses changes in the near-side jet yield to scale long-range non-flow contribution.  In \textsc{ampt}, this leads to rather non-physical results attributable to the significant scattering of final state particles associated with mini-jets. As shown in Figure~\ref{fig:jetyield}, the yield $Y$ of charged particles associated with the near-side jet increases with total event $N_{ch}$. This makes sense, as there is a significant autocorrelation, which is what the technique accounts for. However, shown are results with and without hadronic and partonic scattering, and the two cases differ significantly.   In fact, both the ATLAS and CMS non-flow subtraction methods assume that partons from the jet contribution do not interact in high multiplicity events (i.e., the problem factorizes), but this is not true in \textsc{ampt}.  Thus, any closure test of either method will fail in AMPT, which warrants further studies.

\begin{figure*}[hbtp]
\centering
\includegraphics[width=0.8\linewidth]{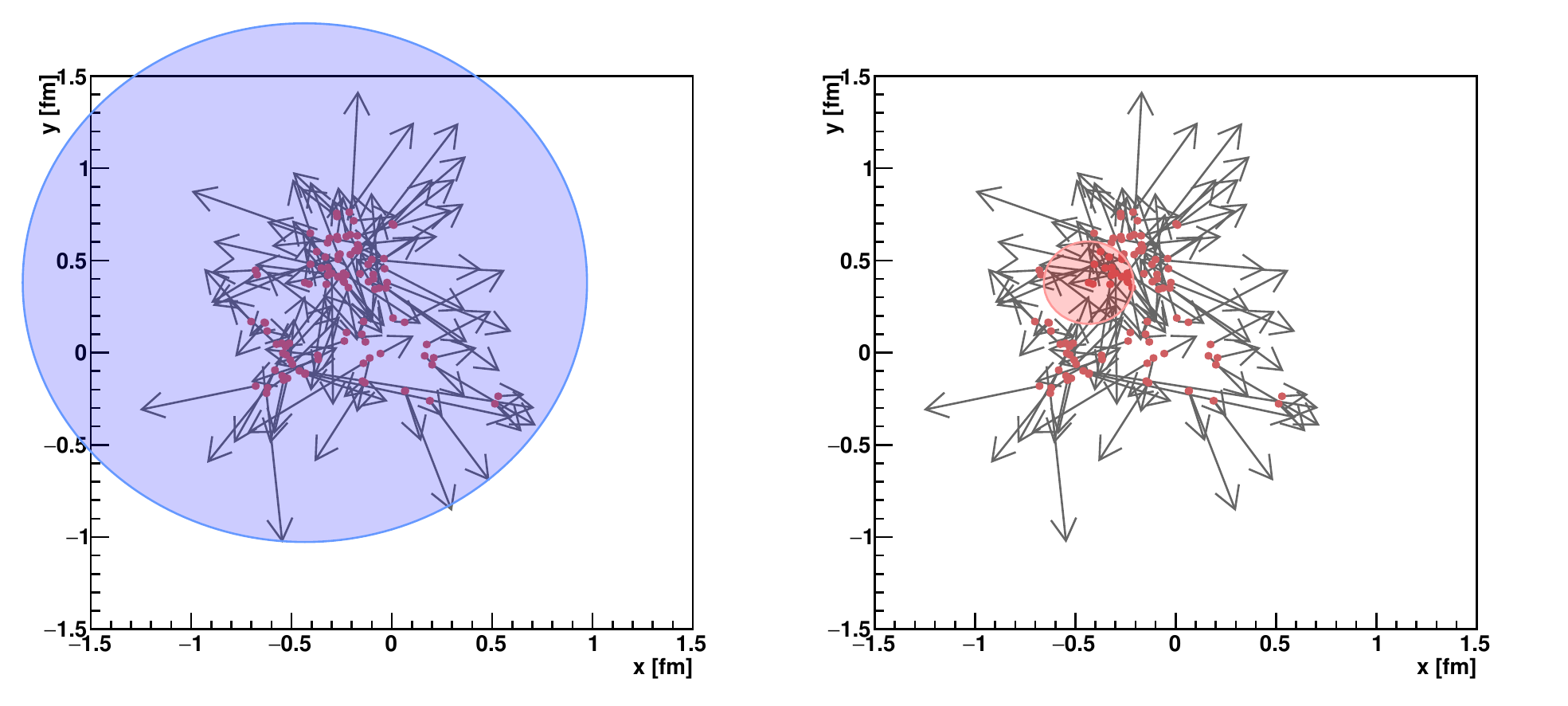}
\caption{A single \textsc{ampt} $p+p$ event with initial partons shown.   The circles correspond to the formation length based on the de Broglie wavelength (left) and the uncertainty principle (right).}
\label{fig:debroglie}
\end{figure*}

The \textsc{ampt} model is, without question, a very useful tool. Nevertheless, it relies on a number of parameters and assumptions that are not physically well motivated. For instance, the formation time of partons corresponds to approximately one tenth of their de Broglie wavelength. Within kinetic theory, this is a surprisingly short amount of time, and simply follows from a basic  uncertainty principle argument.
As illustration of why this matters, Figure~\ref{fig:debroglie} shows the initial partons in a single $p+p$ event. A blue circle (left) is drawn around a given parton of $p_T\approx 0.8$ GeV$/c$, with a radius equal to its de Broglie wavelength $\lambda_B = h/p$; a red circle (right) is drawn for the same parton, with a radius dictated by the uncertainty principle $r = \hbar/p$, thus corresponding to an uncertainty of 100\% on the parton's momentum.  In either case it is clear that multiple other partons lie within the circles, and thus treating the partons as free streaming during their formation time is unlikely to capture the correct physics. In fact, we find that, in these $p+p$ collisions, the $v_n$ coefficients are very sensitive and inversely proportional to the formation time used in the model.

Following feedback received at the conference, we verified that the dilution parameter $D$, equal to the interparticle spacing $\ell$ divided by $\lambda_{\text{mfp}}$, has values $D\ge 1$ at early times in \textsc{ampt}. The Boltzmann equation only holds in the limit  $D << 1$~\cite{Gombeaud:2007ub}, raising questions about the applicability of the particular parton scattering model of \textsc{ampt}. No clear separation exists between the three length scales mentioned at the start of this proceedings, which is necessary for the Boltzmann equation, treating only $2 \rightarrow 2$ scattering, and being insensitive to issues regarding formation time.   QCD makes this an inherently tough problem.   

JLN and JOK acknowledge funding from the Division
of Nuclear Physics of the U.S. Department of Energy
under Grant No. DE-FG02-00ER41152.   JLN also acknowledges generous support from CEA/IPhT/Saclay.





\bibliographystyle{elsarticle-num}
\bibliography{sample}

\begin{thebibliography}{1}
\expandafter\ifx\csname url\endcsname\relax
  \def\url#1{\texttt{#1}}\fi
\expandafter\ifx\csname urlprefix\endcsname\relax\def\urlprefix{URL }\fi
\expandafter\ifx\csname href\endcsname\relax
  \def\href#1#2{#2} \def\path#1{#1}\fi

\bibitem{Nagle:2018nvi}
J.~L. Nagle, W.~A. Zajc, {Small System Collectivity in Relativistic Hadron and
  Nuclear Collisions}\href {http://arxiv.org/abs/1801.03477}
  {\path{arXiv:1801.03477}}.

\bibitem{Nagle:2017sjv}
J.~L. Nagle, R.~Belmont, K.~Hill, J.~Orjuela~Koop, D.~V. Perepelitsa, P.~Yin,
  Z.-W. Lin, D.~McGlinchey, {Minimal conditions for collectivity in $e^+e^-$
  and $p+p$ collisions}, Phys. Rev. C97~(2) (2018) 024909.
\newblock \href {http://arxiv.org/abs/1707.02307} {\path{arXiv:1707.02307}},
  \href {http://dx.doi.org/10.1103/PhysRevC.97.024909}
  {\path{doi:10.1103/PhysRevC.97.024909}}.

\bibitem{Lin:2004en}
Z.-W. Lin, C.~M. Ko, B.-A. Li, B.~Zhang, S.~Pal, {A Multi-phase transport model
  for relativistic heavy ion collisions}, Phys. Rev. C72 (2005) 064901.
\newblock \href {http://arxiv.org/abs/nucl-th/0411110}
  {\path{arXiv:nucl-th/0411110}}, \href
  {http://dx.doi.org/10.1103/PhysRevC.72.064901}
  {\path{doi:10.1103/PhysRevC.72.064901}}.

\bibitem{He:2015hfa}
L.~He, T.~Edmonds, Z.-W. Lin, F.~Liu, D.~Molnar, F.~Wang, {Anisotropic parton
  escape is the dominant source of azimuthal anisotropy in transport models},
  Phys. Lett. B753 (2016) 506--510.
\newblock \href {http://arxiv.org/abs/1502.05572} {\path{arXiv:1502.05572}},
  \href {http://dx.doi.org/10.1016/j.physletb.2015.12.051}
  {\path{doi:10.1016/j.physletb.2015.12.051}}.

\bibitem{Aaboud:2016yar}
M.~Aaboud, et~al., {Measurements of long-range azimuthal anisotropies and
  associated Fourier coefficients for $pp$ collisions at $\sqrt{s}=5.02$ and
  $13$ TeV and $p$+Pb collisions at $\sqrt{s_{\mathrm{NN}}}=5.02$ TeV with the
  ATLAS detector}, Phys. Rev. C96~(2) (2017) 024908.
\newblock \href {http://arxiv.org/abs/1609.06213} {\path{arXiv:1609.06213}},
  \href {http://dx.doi.org/10.1103/PhysRevC.96.024908}
  {\path{doi:10.1103/PhysRevC.96.024908}}.

\bibitem{Khachatryan:2016txc}
V.~Khachatryan, et~al., {Evidence for collectivity in pp collisions at the
  LHC}, Phys. Lett. B765 (2017) 193--220.
\newblock \href {http://arxiv.org/abs/1606.06198} {\path{arXiv:1606.06198}},
  \href {http://dx.doi.org/10.1016/j.physletb.2016.12.009}
  {\path{doi:10.1016/j.physletb.2016.12.009}}.

\bibitem{Gombeaud:2007ub}
C.~Gombeaud, J.-Y. Ollitrault, {Elliptic flow in transport theory and
  hydrodynamics}, Phys. Rev. C77 (2008) 054904.
\newblock \href {http://arxiv.org/abs/nucl-th/0702075}
  {\path{arXiv:nucl-th/0702075}}, \href
  {http://dx.doi.org/10.1103/PhysRevC.77.054904}
  {\path{doi:10.1103/PhysRevC.77.054904}}.

\end{thebibliography}







\end{document}